\documentclass[twocolumn,fullpage]{revtex4} 
\usepackage{graphicx,subfigure,color,amsmath,amssymb}

\begin{document}

 \title{Exact results for a simple epidemic model on a directed network: Explorations of a 
system in a non-equilibrium steady state.}
 \author{Maxim S.~Shkarayev$^{1}$}
 \author{R. K. P.~Zia$^{1,2}$}
 \affiliation{$^{1}$ Department of Physics \& Astronomy, Iowa State University, Ames, IA, 
50011\\
 $^{2}$ Physics Department, Virginia Polytechnic Institute and State University, 
Blacksburg, VA, 24061}

 \begin{abstract}
 Motivated by fundamental issues in non-equilibrium statistical mechanics (NESM), we study 
the venerable susceptible-infected (SIS) model of disease spreading in an idealized, simple 
setting. Using Monte Carlo and analytic techniques, we consider a fully connected, 
uni-directional network of odd number of nodes, each having an equal number of in- and 
out-degrees. With the standard SIS dynamics at high infection rates, this system settles 
into an active non-equilibrium steady state. We find the exact probability distribution and 
explore its implications for NESM, such as the presence of persistent probability currents.
 \end{abstract}

\maketitle

\section{Introduction}

 Nearly all interesting phenomena around us are non-equilibrium stochastic processes, from 
all forms of living organisms to the life-sustaining atmosphere and sun. Yet, very little 
is understood about non-equilibrium statistical systems, especially in comparison to the 
highly successful Boltzmann-Gibbs framework for systems in thermal equilibrium. Of course, 
the most important distinction between the two is that, for the latter, once the energy 
functional (Hamiltonian $\mathcal{H}$) of the system and the properties of the reservoirs 
(e.g., temperature $T$, chemical potential $\mu $) are specified, the time-independent 
microscopic probability distribution, $P^{\ast}$, is known (e.g., a Boltzmann factor, 
$P^{\ast}\propto e^{-\mathcal{H}/k_{B}T}$). Furthermore, if time-dependent behavior is to 
be modeled for such systems, a stochastic dynamics can be readily written down, following 
the rule of detailed balance. One physical consequence for such equilibrium states
is that there are no \textit{net} 
exchanges (e.g., of energy, particles) between the system and its reservoirs. By 
contrast, we may wish to describe a system in contact with many reservoirs so that, even 
when it is in a steady (stationary) state, non-trivial net exchanges of various 
quantities exist. In other words, there are typically net fluxes \textit{through} such systems, 
as they settle into non-equilibrium steady states (NESS). No one doubts that the existence 
of our ecosystem depends crucially on such a steady flux of radiant energy, from the sun 
and to the outer-space. Now, to describe such systems, we must use dynamical rules which 
violate detailed balance or time-reversal. Then, we face many serious challenges, perhaps 
the simplest being the following. Given a set of detailed balance violating, stochastic 
rules of evolution, the system will settle into a NESS; but what is the associated 
stationary probability distribution, $P^{\ast}$? In addition, it is not surprising that, in 
analogy with magnetostatics, there will be non-trivial steady (probability) currents, 
$K^{\ast}$, with which the average net fluxes of observables can be 
computed~\cite{ZS07,ZS06}. Although a method for constructing $P^{\ast}$ and $K^{\ast}$ is 
known~\cite{Hill66,Schn76}, it is formal and quite cumbersome. As a result, computing 
observables with them is hopelessly difficult, while the physics behind these expressions is 
far from discernible. In particular, there are very few systems for which analytic forms for 
$P^{\ast}$ and $K^{\ast}$ are known explicitly. In this context, we study simple model 
systems -- motivated primarily by natural phenomena -- which settle into non-trivial NESS, 
with the goal of gaining some insight into the issues presented above.

In this paper, we consider the venerable SIS model of 
epidemics~\cite{KermackMcKendrick,AndersonMay,DaleyGani,allen2008}, in which an individual 
of a population can be in an infected ($I$) or a susceptible ($S$) state. While an $I$ 
spontaneously recovers with some rate, an $S$ can become infected, depending on its 
connectivity to others and their conditions. If the ratio of infection-to-recovery rates is 
high enough, a finite fraction of the population are $I$'s, an `epidemic' is present, and 
the system is said to be in an `active' state. In the simplest model, as soon as all $I$'s 
have recovered, there will be no further evolution, a state labeled as `inactive.' Of course 
in reality, spontaneous reinfections (i.e., not due to another $I$) do occur and the 
inactive state may be characterized as having a vanishingly small fraction of $I$'s on the 
average. For public health organizations, the transition between inactive and active states 
is clearly of major concern. Our interest here is more theoretical, namely, when is an 
active state a NESS and what are its novel characteristics. In particular, in most model 
studies, $\tau_{j}^{i}$, the probability an infected individual $i$ can affect a 
susceptible $j$, is the same as $\tau_{i}^{j}$. In reality, 
infection rates are typically asymmetric (due to,
e.g., inherently different immune systems or different habits of personal
hygiene), leading us to expect the active states to be NESS. While an
undirected graph is adequate for describing the network in the symmetric
case, digraphs (i.e., directed graphs) will be needed for a system with 
$\tau _{j}^{i}\neq \tau _{i}^{j}$. Our goal here is to explore systems which
not only lead to prominently observable effects of detailed balance
violation, but also are on the same footing as models obeying detailed
balance. These models allow us to construct quantitative and meaningful
comparisons between equilibrium and non-equilibrium stationary states. As
will be shown, it is remarkable (and fortunate) that we are able to find the
explicit analytic forms for $P^{\ast }$ and $K^{\ast }$, for an NESS of a
well-mixed SIS system with asymmetric infection probabilities, deep in the
active phase.

 The rest of this paper is organized as follows. In the next section, we present a detail 
description of the model. Section III will be devoted to the master equation governing the 
evolution of the probability distribution and a discussion of the role of detailed balance 
in the dynamics. The exact, microscopic stationary 
probability distribution and the associated steady 
currents are provided in a following section. The observable consequences of the 
underlying persistent currents are explored, with the introduction of a novel macroscopic 
quantity. After a section on simulation results, we conclude with a summary and outlook. 
Some technical details are provided in Appendices.

\section{Model Specifications}

\label{sec:model}

 We consider the simplest of SIS 
models~\cite{KermackMcKendrick,AndersonMay,DaleyGani,allen2008} on fully connected networks 
of $N$ nodes, evolving stochastically according to the following rules. For reasons to be 
made clear, we restrict ourselves to odd $N$ ($=2\ell +1$). A node, labeled by $i$ 
($=1,...,N$), can be found in one of two states: $I$ or $S$, 
infected with or susceptible to a 
disease, respectively. We specify a configuration (microstate) of the system by $\vec{m}$,
(a vector) with entries $m_{i}=0$ or $1$, when node $i$ is susceptible or infected, 
respectively. Thus,
 \begin{align}
 &n\left(\vec{m}\right) \equiv \Sigma_{i}m_{i}
 \end{align}
 is the number of infected individuals in microstate $\vec{m}$. Now, configuration space 
consists of the vertices of a unit cube in $N$ dimensions, while the evolution of our 
system corresponds to moving from vertex to vertex, only along an edge of this cube. 
Specifically, the changes occur at discrete time steps, with exactly one event taking 
place: Either an infected node becomes susceptible, or vice versa. In the language of a 
kinetic Ising model, these moves correspond to Glauber spin-flip dynamics~\cite{Glauber63}. 
Note that our system will always change its state in a step, though the ratio of recovery 
to infected processes differ in general. Though such a rule seems artificial, it is in the 
spirit of the well-established Gillespie algorithm in Monte Carlo 
simulations~\cite{Gill77}. When performing computer simulations, taking $N$ steps is 
referred to as a sweep or a Monte Carlo step (MCS), in which period every node has, on the 
average, one chance to change its state.

 In our model, the recovery process occurs with probability proportional to 
$r>0$. Meanwhile, an infected node $j$ can transmit the disease to a susceptible node 
$i$ with probability proportional to $\tau_{i}^{j}$, where the following class of 
$\tau_i^j$'s is considered. Since we do not allow a node to infect itself, we impose 
$\tau _{i}^{i}\equiv 0$. As we wish to consider possibly \textit{asymmetric} 
infection probabilities $\tau _{i}^{j}\neq \tau _{j}^{i}$, a convenient way to encode 
this information is (for $i\neq j$)
 \begin{align}
 &\tau _{i}^{j}=\theta \left( 1+\sigma a_{i}^{j}\right) ,  \label{theta}
 \end{align}
 where $\theta $ controls the overall rate of infection, $\sigma $ is a
parameter in the interval $\left[ 0,1\right] $, and $a_{i}^{j}$ is a
skew-symmetric matrix with elements $\pm 1$. The advantage of the form (\ref{theta}) is that the symmetric and \textit{antisymmetric} aspects of the
infection rates are shown explicitly, controlled by $\theta $ and $\theta
\sigma $, respectively. Thus, if $\tau _{i}^{j}=0$ ($i$ cannot be infected
by $j$), then $i$ can infect $j$ with probability $2\theta $. Furthermore, $\sigma $ allows us to tune continuously, from an ordinary SIS model to a
network with \textit{maximally asymmetric} infection rates.

Given that $N$ is odd, we can impose $\sum_{j}a_{i}^{j}=0$, a condition
which means that, for $\sigma =1$, each node will have precisely $\ell $ in-
and out-degrees. In other words, in this special network, each individual
can be infected by half of the (rest of the) population and \textit{immune}
to the other half. From the form of (\ref{theta}) and $\Sigma _{j}a_{i}^{j}=0
$, it may be argued that the `average rate' for an individual to be infected
is controlled only by $\theta $, so that it is meaningful for us to compare
the epidemics levels in networks with different $\sigma $'s. 

Obviously, for $\sigma =0$, the infection rate are symmetric and uniform,
representing the venerable `well-mixed' SIS model~\cite{allen2008}. In the
large $N$ limit, this should have vanishingly small fluctuations, and, with
no spatial structure, it can be well described by deterministic the rate
equation~\cite{allen2008}: $dn/dt=-rn+\theta n(N-n)$. As $t\rightarrow
\infty $, $n\left( t\right) $ will settle into one of two fixed points ($n^{\ast }$): an `inactive' state ($n^{\ast }=0$), if the infection rate is
too low, or an `active' one with $n^{\ast }=N-r/\theta $. The transition
occurs at critical ratio $\left( \theta /r\right) _{c}=1/N$. There is also
much known about such a model on other networks, e.g., those corresponding
to populations with spatial structure \cite{Moore2000,Pastor-Satorras2001,Keeling2005,Barrat2008}. 

As it stands, the unique stationary state in the stochastic version is an
absorbing state ($\vec{m}=\vec{0}$). For large/small infection/recovery
rates, this state is rarely reached and\ the active state is referred to as
quasistationary. We choose a different rule, so that a non-trivial, active
state exists as genuinely stationary, namely, by infecting a randomly chosen
node whenever the system arrives at $\vec{m}=\vec{0}$. While such a rule
will affect the precise determination of the critical parameters for the
transition between inactive to active states, it should not play a serious
role for systems far in the active state. Should we extend our studies to
the critical region, we can always modify this rule to reinfect this state
with an arbitrarily small probability.

Let us emphasize that the model presented here is highly specialized,
designed to highlight the differences between equilibrium states and NESS,
rather than to describe a realistic population. Neverthless, our main result
-- the presence of cyclic behavior and its quantitative characterization --
is expected to prevail in all epidemics, even though these effects are not
likely to be dominant.

\section{Master equation and detailed balance}\label{sec:MEDB}

 The full stochastic process specified above is described by a master equation for 
$P\left(\vec{m},t\right) $, the probability for finding our system in microstate $\vec{m}$, 
$t$ steps from some initial configuration. (Since our focus will be the stationary state, 
reached after very long times, the initial state is irrelevant and will not be explicitly 
shown here.) In general, the master equation reads
 \begin{align}
 &P\left(\vec{m},t+1\right) =\sum_{\vec{m}^{\prime}}R\left(\vec{m}\leftarrow 
\vec{m}^{\prime}\right) P\left(\vec{m}^{\prime},t\right),
 \label{P=LP}
 \end{align}
 where $R$ represents the transition probability, to go from $\vec{m}^{\prime}$ to 
$\vec{m}$. In our model, these $\vec{m}$'s differ by only one entry (e.g., 
$m_{i}^{\prime}=1-m_{i}$), so that we can simplify the above to:
 \begin{align}
 \begin{split}
 &P\left(m_{1},...,m_{N},t+1\right) =\\
 &\sum_{i}\Omega_{i}\left(\vec{m}^{\prime}\right) 
P\left(m_{1},...,1-m_{i},....m_{N},t\right), \label{P=OmP}
 \end{split}
 \end{align}
 where $\Omega$ are the transition probabilities spelled out above. To find the explicit 
expressions, consider first $S$ to $I$ transitions. For simplicity, 
we define our model~\footnote{There are many other ways to introduce infection from 
two or more individuals. For example, if each can infect our $i$ independently with 
probability $\lambda$, then we would write $\gamma = 1 - (1-\lambda)^n$ instead.} 
by letting 
 \begin{align}
 &\gamma_{i}\left(\vec{m}\right) \equiv \sum_{j}\tau_{i}^{j}m_{j}.
 \label{gamma1}
 \end{align} be the rate the node $i$ in $\vec{m}$ becomes infected. 
 Here, the $m_{j}$ insures that $j$ is infected, while 
$\tau_{i}^{j}$ embodies both the infection probability and the connectivity between $i$ and 
$j$. Substituting~(\ref{theta}), we see that
 \begin{align}
 &\gamma_{i}\left(\vec{m}\right) =\theta \left[ n\left(\vec{m}\right)
 +\sigma \kappa_{i}\left(\vec{m}\right) \right], \label{gamma2}
 \end{align}
 where
 \begin{align}
 &\kappa_{i}\left(\vec{m}\right) \equiv \sum_{j}a_{i}^{j}m_{j} \label{kappa}
 \end{align}
 represents an \textit{excess} of the infected individuals who can affect $i$, over those 
which cannot do so. On the average, $\kappa $ would be zero, as $a_{i}^{j}$ assigns $+1$ to 
infected individuals with a link directed to $i$ and $-1$ to ones direct away from $i$. 
Thus, the total rate for \textit{any} susceptible individual in $\vec{m}$ to be infected is 
(proportional to) the sum
 \begin{align}
 \begin{split}
 &\sum_{i}\left(1-m_{i}\right) \gamma_{i}\left(\vec{m}\right) =\theta
 \sum_{ij}\left(1-m_{i}\right) \left(1+\sigma a_{i}^{j}\right) m_{j} \\
 &=n\left(\vec{m}\right) \left[ N-n\left(\vec{m}\right) \right] \theta,
 \label{SumGam2}
 \end{split}
 \end{align}
 where the last equality arises from 
$a_{i}^{j}=-a_{j}^{i}$ and $\Sigma_{i}a_{i}^{j}=0$. The 
significance of this class of asymmetric networks is 
revealed: The \textit{total} rate of infection does not 
depend on the details of the digraph (i.e., $a_{j}^{i}$).

 For $I$ to $S$ transitions, since each of the $n$ infected nodes can recover independently, 
the total recovery rate is (proportional to) $rn\left(\vec{m}\right)$. These results 
provide us with the normalization factor
 \begin{align}
 &\rho \left(\vec{m}\right) =\frac{1}{n\left(\vec{m}\right) r+n\left(\vec{m}\right) \left[ 
N-n\left(\vec{m}\right) \right] \theta}.
 \end{align}
 Note that this factor depends on $\vec{m}$ only through $n$, the total number of infected, 
rather than the details of each individual. Thus, whenever there is no confusion, we will 
use the simpler notation
 \begin{align}
 &\rho_{n}=\frac{1}{nr+n\left(N-n\right) \theta}.
 \end{align}
 Of course, this expression is singular for $\vec{m}=0$, a special case for which 
$\Omega_{i}\left(\vec{0}\right) $ is simply $1/N$.

 With these forms, $\Omega_{i}\left(\vec{m}\right) $ is explicitly $\rho 
\left(\vec{m}\right) \left[ m_{i}r+\left(1-m_{i}\right) \gamma_{i}\left( \vec{m}\right) 
\right] $, for $\vec{m}\neq 0$. Since $m$ can be either $0$ or $1$, $\Omega $ is given by 
one or the other term here. Note that, for the fully infected state, $\Omega_{i}$ reduces 
to $\rho_{N}r=1/N$ for all $i$, which is completely consistent with our expectations. Inserting 
these $\Omega $'s into Eqn.~(\ref{P=OmP}), we have the full master equation. One subtlety 
we should emphasize is that, in Equation~(\ref{P=OmP}) the argument in $\Omega_{i}$ is 
$\vec{m}^{\prime}$, which is $\vec{m}$ \textit{except}\ for entry $i$, while their $n$'s 
differs by unity. Thus, it is worthwhile writing the master equation explicitly
 \begin{align}
 \begin{split}
 &P\left(m_{1},...,m_{N},t+1\right) =\\
 &=\sum_{i}\left\{\rho_{n+1}\left(1-m_{i}\right) 
r+\rho_{n-1}m_{i}\gamma_{i}\left(\vec{m}^{\prime}\right)\right\} \times\\
 &\times P\left(m_{1},...,1-m_{i},....m_{N},t\right), \label{ME}
 \end{split}
 \end{align}
 where $n=n\left(\vec{m}\right) \in \left[ 2,N-1\right]$. Note that the two terms in 
$\left\{...\right\} $ correspond to recovery and infection, respectively. Of course, these 
terms must be suitably modified for $n=1$ and $N$.

 Given a set of transition probabilities, it is simple to see if they obey detailed balance 
using the Kolmogorov criterion~\cite{Kolmo36}. Consider a closed loop involving $L$ 
configurations,
 $\vec{m}^{\left(1\right)}\rightarrow \vec{m}^{\left(2\right)}\rightarrow \ldots 
\rightarrow \vec{m}^{\left(L\right)}\rightarrow \vec{m}^{\left(1\right)}$,
 as well as the product
 $$R\left(\vec{m}^{\left(1\right)}\leftarrow \vec{m}^{\left(L\right)}\right) \ldots 
R\left(\vec{m}^{\left(3\right)}\leftarrow \vec{m}^{\left(2\right)}\right) R\left( 
\vec{m}^{\left(2\right)}\leftarrow \vec{m}^{\left(1\right)}\right) $$
 along it and the product
 $$R\left(\vec{m}^{\left(1\right)}\leftarrow \vec{m}^{\left(2\right)}\right) \ldots 
R\left(\vec{m}^{\left(L-1\right)}\leftarrow \vec{m}^{\left(L\right)}\right)R\left( 
\vec{m}^{\left(L\right)}\leftarrow \vec{m}^{\left(1\right)}\right)$$
 for traversing the loop in reverse. If and only if these products are equal for 
\textit{all} loops, detailed balance is satisfied. Then, the stationary distribution can be 
thought of as one in thermal equilibrium, with no net probability currents anywhere. In 
Appendix~\ref{DB+KC}, we provide some details which show that, in general, detailed balance 
is violated if $\sigma >0$. It is hardly surprising that an SIS model on a complete, 
undirected graph settles into an equilibrium state (with zero net currents, as in 
electrostatics). By contrast, systems with $\sigma >0$ will evolve towards 
\textit{non-}equilibrium steady states with persistent currents (as in 
magnetostatics)~\cite{ZS07}. One of the goals of this study is to show, both analytically 
and in Monte Carlo simulations, the existence of these currents and their implications for 
observables. But first, let us find the stationary distribution.

\section{Exact steady state distribution and persistent probability 
currents}\label{sec:exact}

It is well known that the equation~(\ref{P=LP}), with the transition probabilities given 
here, will evolve $P$ to a stationary state, which we denote by $P^{\ast}\left( 
\vec{m}\right) $. If the dynamics satisfies detailed balance, then finding this $P^{\ast}$ 
is a trivial process. Otherwise, though there is a systematic method to construct 
$P^{\ast}$~\cite{Hill66}, this route is prohibitively cumbersome and, typically, finding an 
explicit $P^{\ast}$ is essentially impossible. Nevertheless, under a few special 
circumstance, such $P^{\ast}$'s have been found. The simplest example is biased diffusion 
on a ring. Introduced as the asymmetric exclusion process~\cite{Liggett,Spohn,Schutz}, 
$P^{\ast}\propto 1$ was known long ago~\cite{Spitzer70}. Here, we are able to find a 
non-trivial $P^{\ast}$, based on an Ansatz inspired by simulation results. Since our 
dynamics is clearly ergodic, this $P^{\ast}$ is unique, so it is \textit{the} stationary 
distribution.

\subsection{$P^{\ast}$ for the undirected network}

Before we present the general result, let us recapitulate well-known results, for the 
reader's convenience, in the simple SIS model on a complete and undirected network ($\sigma 
=0$). Of course, due to the reinfection of the inactive state, our results for $P^{\ast}$ 
are slightly different from the distribution for a quasistationary state.

 Since this dynamics satisfies detailed balance, we simply start with an unknown 
$P^{\ast}\left(\vec{0}\right) $ and obtain the rest by repeated use of 
$R\left(\vec{m}\leftarrow \vec{m}^{\prime}\right) P^{\ast}\left(\vec{m}^{\prime}\right) 
=R\left(\vec{m}^{\prime}\leftarrow \vec{m}\right)P^{\ast}\left(\vec{m}\right) $. Note that 
this condition reflects the simple balance between the infection and recovery rates for any 
\textit{single} individual. For a complete, undirected graph, it is clear that 
$R\left(\vec{m}^{\prime}\leftarrow \vec{m}\right) \ $depends only on $n$. Therefore, 
$P^{\ast}\left(\vec{m}\right) $ is also a function of $n\left(\vec{m}\right) $ only and so, 
we write:
 \begin{align}
 &P^{\ast}\left(\vec{m}\right) =P_{n}.  \label{Pm=Pn}
 \end{align}
 In terms of these, the balance of the rates for a single node (i.e., $\rho_{n}rP_{n}$ for 
recovery and $\rho_{n-1}\left(n-1\right) \theta P_{n-1}$ for infection) leads to
 \begin{align}
 &\rho_{n}rP_{n}=\rho_{n-1}\left(n-1\right) \theta P_{n-1}.  \label{RR}
 \end{align}
 This recursion allows us to express the $P_{n}$'s in terms of $P_{0}$, starting with the 
special case $\rho_{1}rP_{1}=P_{0}/N$. Thus,
 \begin{align}
 \begin{split}
 &P_{n}=\left(n-1\right) !\alpha ^{n-1}\frac{P_{0}}{\rho_{n}rN}\label{ExactSol}= \\
 &=\left\{\phi_{n}\alpha ^{n-1}\left(n-1\right) !+\left(1-\phi_{n}\right) \alpha 
^{n}n!\right\}P_0
 \end{split}
 \end{align}
 for $n\ge 1$, where
 \begin{align}
 &\phi_{n}\equiv n/N
 \end{align}
 is the fraction (of the infected in $\vec{m}$), and
 \begin{align}
 &\alpha \equiv \theta /r
 \end{align}
 is the ratio of the rates (which is clearly the only quantity of significance here). We 
remark that the various factors in Eqn.~(\ref{ExactSol}) lend themselves to intuitive 
interpretations: relative weights for the infected and susceptible factions, cumulative 
factors for infection ($\alpha ^{n}$), and combinatorics.

 Finally, the unknown $P_{0}$ can be fixed by imposing normalization, namely, 
$1=\sum_{\vec{m}}P^{\ast}\left(\vec{m}\right) =\sum_{n=0}^{N}\binom{N}{n}P_{n}$ (to account 
for the $\binom{N}{n}$ microstates $\vec{m}$ for a specific $n$). Thus,
 \begin{align}
 \begin{split}
 &\frac{1}{P_{0}}= 1+\sum_{n=1}^{N}\binom{N}{n}\frac{n!\alpha ^{n-1}}{N}+\\
 &+ \sum_{n=1}^{N}\binom{N}{n}\frac{n!\alpha ^{n}}{N}\left(N-n\right).\label{P0}
 \end{split}
 \end{align}
 In Appendix~\ref{norm}, we show $P_{0}$ can be expressed compactly as
 \begin{align}
 &P_{0}=\frac{e^{-1/\alpha}}{2\alpha ^{N-1}\Gamma \left(N,1/\alpha \right)}.
 \label{P0-last}
 \end{align}
 where $\Gamma $ is an upper incomplete gamma function. Note that it may appear 
counter-intuitive that, by setting the infection rate $\theta $ to zero, $P_{0}=1/2$ is 
less than unity. This result is merely an artifact of our special rule for reinfecting the 
absorbing state as soon as it is reached. If this rule is modified appropriately, $P_{0}$ 
can be made arbitrarily close to unity.

\subsection{$P^{\ast}$ for an asymmetric network}

 Let us turn to the general $\sigma >0$ case, in which $i$ can infect $j$ with a rate 
different from the opposite situation. Though there is no \textit{a priori} reason to 
expect $P^{\ast}\left(\vec{m}\right) $ to depend only on $n\left(\vec{m}\right) $, we are 
inspired by simulation results (shown below) indicating that this property persists. Thus, 
we attempt to find a stationary solution to Eqn.~(\ref{P=OmP},\ref{ME}) with $\sigma >0$ by 
using an \textit{Ansatz}: $P^{\ast}\left(\vec{m}\right) =\tilde{P}_{n}$. Substituting this 
Ansatz into
 \begin{align}
 \begin{split}
 &\tilde{P}_{n}=P^{\ast}\left(m_{1},...,m_{N}\right) =\\
 &\sum_{i}\left\{\rho_{n+1}\left(1-m_{i}\right) 
r+\rho_{n-1}m_{i}\gamma_{i}\left(\vec{m}^{\prime}\right) \right\} \times\\
 &\times P^{\ast}\left(m_{1},...,1-m_{i},....m_{N}\right),\label{PtildeME}
 \end{split}
 \end{align}
 we see that the right hand side reduces to the following two terms:
 \begin{align}
 &\sum_{i}\rho_{n+1}\left(1-m_{i}\right)r\tilde{P}_{n+1}+
 \sum_{i}\rho_{n-1}m_{i}\gamma_{i}\left(\vec{m}^{\prime}\right)\tilde{P}_{n-1}.
 \end{align}
 Since $\tilde{P}_{n+1}$ does not depend on $i$, the first sum leads to 
$\rho_{n+1}\left[N-n\right] r\tilde{P}_{n+1}$. To carry out the sum in the second 
requires a little more care, since $\vec{m}^{\prime}$ stands for 
$\left(m_{1},...,1-m_{i},....m_{N}\right) $, with 
$n\left(\vec{m}^{\prime}\right)=n-1$. Thus,
 \begin{align}
 &\sum_{i}m_{i}\gamma_{i}\left(\vec{m}^{\prime}\right) = \theta 
\sum_{i}m_{i}\left[n-1+\sigma \Sigma_{j}a_{i}^{j}m_{j}\right] =n\left(n-1\right) \theta
 \end{align}
 is \textit{independent} of $\sigma $, while~(\ref{PtildeME}) becomes
 \begin{align}
 &\tilde{P}_{n}=\rho_{n+1}\left[ N-n\right] r\tilde{P}_{n+1}+\rho_{n-1}n\left(n-1\right) 
\theta \tilde{P}_{n-1}, \label{Ptilde}
 \end{align}
 for $n\geq 2$. Since our reinfection rule for $\vec{m}=0$ is special, we need to 
supplement 
these 
with
 \begin{align}
 & \tilde{P}_{1}=\rho_{2}\left[ N-1\right] r\tilde{P}_{2}+\left(1/N\right) \tilde{P}_{0}.
 \end{align}

 A solution to this set of equations can be found directly; however, given that 
they are independent of $\sigma $, it behooves us to consider $\tilde{P}_{n}=P_{n}$. 
Recalling $\rho_{n}rN=1/\left\{\phi_{n}+\left(1-\phi_{n}\right) n\alpha \right\} $, it is 
straightforward to check that
 \begin{align}
 \rho_{n+1}\left[ N-n\right] r\frac{P_{n+1}}{P_{n}} =\frac{\left(1-\phi_{n}\right) \alpha 
n}{\phi_{n}+\left(1-\phi_{n}\right) n\alpha} &\\
 \rho_{n-1}n\left(n-1\right) \theta \frac{P_{n-1}}{P_{n}} 
=\frac{\phi_{n}}{\phi_{n}+\left(1-\phi_{n}\right) n\alpha}&,
 \end{align}
 and so, $P_{n}$ indeed satisfies Eqn.~(\ref{Ptilde}). Thus, our expectation, that 
expression~(\ref{ExactSol}) is the stationary distribution $P^{\ast}\left(\vec{m}\right) $ 
for \textit{any} $\sigma $, is verified.

\subsection{Persistent probability currents and their consequences}

 Since Eqn.~(\ref{P=LP}) is a continuity equation for the probability density, it is 
natural to regard the right hand side as a sum over probability currents. In our case, the 
\textit{net} current from microstate $\vec{m}^{\prime}$ to $\vec{m}$ (over the single time 
step $t\rightarrow t+1$) can be identified as
 \begin{align}
 \begin{split}
 &K\left(\vec{m}^{\prime}\rightarrow \vec{m},t\right) =\\
 &=R\left(\vec{m}\leftarrow\vec{m}^{\prime}\right) P\left(\vec{m}^{\prime},t\right)-R\left( 
\vec{m}^{\prime}\leftarrow \vec{m}\right) P\left(\vec{m},t\right).
 \end{split}
 \end{align}
 In the steady state, we denote this quantity by
 \begin{align}
 \begin{split}
 &K^{\ast}\left(\vec{m}^{\prime}\rightarrow \vec{m}\right) =\\
 &=R\left(\vec{m}\leftarrow \vec{m}^{\prime}\right) 
P^{\ast}\left(\vec{m}^{\prime}\right)-R\left(\vec{m}^{\prime}\leftarrow \vec{m}\right) 
P^{\ast}\left(\vec{m}\right).
 \end{split}
 \end{align}
 Thus, if the underlying dynamics obeys detailed balance, $K^{\ast}$ vanishes everywhere. 
Otherwise, there must be non-trivial $K^{\ast}$'s, which we refer to as \textit{persistent 
currents}. Since only one individual can change state, a current is naturally associated 
with an edge of the $N$-cube. For example, for $m_{i}=1\leftrightarrow m_{i}^{\prime}=0$, 
this current is
 \begin{align}
 &K^{\ast}\left(\vec{m}^{\prime}\rightarrow \vec{m}\right) 
=\rho_{n}\gamma_{i}\left(\vec{m}^{\prime}\right) P_{n}-\rho_{n+1}rP_{n+1},
 \end{align}
 where $n$ stands for $n\left(\vec{m}^{\prime}\right) $. Using 
Eqns.~(\ref{gamma2},\ref{RR}), we have
 \begin{align}
 \begin{split}
 &K^{\ast}\left(\vec{m}^{\prime}\rightarrow \vec{m}\right) =\rho_{n}\theta \left[ n+\sigma 
\kappa_{i}\left(\vec{m}\right) \right]P_{n}-\rho_{n+1}rP_{n+1}= \\
 &=\sigma \kappa_{i}\left(\vec{m}^{\prime}\right) \rho_{n}\theta P_{n},
 \end{split}
 \end{align}
 (apart from the cases near $n=0,N$). From~(\ref{kappa}), we see that 
$\kappa_{i}\left(\vec{m}\right) =\kappa_{i}\left(\vec{m}^{\prime}\right) $ since both are 
independent of $m_{i}$. The final expression is
 \begin{align}
 K^{\ast}\left(\vec{m}^{\prime}\rightarrow \vec{m}\right) =\left[ 
\sigma\kappa_{i}\left(\vec{m}\right) \right] \alpha ^{n}\left(n-1\right)!P_{0}/N 
\label{K*},
 \end{align}
 showing explicitly that it vanishes with the product of the asymmetry strength ($\sigma$)
 and the `excess' ($\kappa_{i}$) of infected individuals connected to $i$.

Such microscopic currents (on a discrete space) are analogous to current densities in
electrodynamics and hydrodynamics. In a stationary state, the divergence free condition 
implies that the K*'s must form closed loops. In analogy with fluid 
dynamics, we may refer to the `curl' 
of such (probability) currents as `probability vorticity,' $\omega $. In our discrete 
configuration space, such an $\omega $ should be associated with a face of the cube 
(plaquette) and defined as the \textit{sum} of the currents around the face 
($\thicksim \oint \vec{j}\cdot d\vec{\ell}$ in hydrodynamics). Let us 
consider the vorticity around the $i$-$j$ plaquette (i.e.,~\ref{DBKC1}). The four 
$K^{\ast}$'s involved starts with the state 
$\left\{m_{i},m_{j},\hat{m}\right\}=\left\{0,0,\hat{m}\right\} $, where $\hat{m}$ denotes 
$m_{k\neq i,j}$, with $\nu $ infected individuals. With details shown in 
Appendix~\ref{vorticity}, this current loop sums to 
 \begin{align}
 \begin{split}
 &\omega_{ij}^{\ast}=\left(P_{0}/N\right) \alpha^{\nu}\left(\nu -1\right)! \times\\
 &\times\left[(\alpha \nu -1)\sum_{k\neq i,j}(a_{j}^{k}-a_{i}^{k})m_{k} + 2a_{j}^{i}\right].
 \label{omega}
 \end{split}
 \end{align}
 If we sum over all possible $\hat{m}$'s, all details of the rest of the system (such as 
$\nu $) disappear and the result can be regarded as a `coarse-grained' vorticity:
 \begin{align}
 &\omega_{ij|cg}^{\ast}\equiv \sum_{\left\{\hat{m}\right\}}\omega_{ij}^{\ast}.
 \end{align}
 Not surprisingly, such a vorticity is 
proportional to the key ingredients of asymmetry, $\sigma a_{j}^{i}$: 
 \begin{align}
 \omega _{ij|cg}^{\ast }=\hat{\omega}a_{j}^{i},  \label{omega-cg}
 \end{align}
 where
 \begin{align}
 \hat{\omega}=\frac{\sigma}{NS_{N-1}\left(\alpha\right)}
 \left(\frac{S_{N-2}(\alpha)-1}{N-2}+\alpha S_{N-3}(\alpha )\right) \label{eq:omegahat}
 \end{align}
 depends on, apart from $\sigma$, only the basic control parameters $N$ and $\alpha$. Here, $S_N(\alpha)$ is 
defined in Eqn.~(\ref{eq:SN}).

While the analysis above is valuable at the microscopic level, the behavior of macroscopic 
observables are often more interesting, in that they exemplify collective behavior in a 
statistical mechanical system. For example, though the microscopic distribution of an Ising 
model is trivially analytic, the properties of the total mangetisation (analog of $n$ here) 
signal phase transitions and display highly non-trivial singularities. In this spirit, we 
turn to macroscopic observables which reveal the presence of probability current loops. In 
classical mechanics, mass currents (and loops) are ubiquitous. For example, in rotation of 
rigid bodies, these currents are more commonly characterized by the total angular momentum 
$\vec{L}=\int \vec{r}\times \vec{v}\rho \left(\vec{r}\right) dr$. Exploiting the notion 
that $\rho \vec{v}$ represents the mass current, we will introduce the analog of $\vec{L}$\ 
here, in the context of the simplest of examples.

 Consider two subgroups of our population, labeled by $g=1,2$. For convenience, let them 
have equal size: $N_{1}=N_{2}$. An obvious macroscopic variable is the pair 
$\left(n_{1},n_{2}\right) $, the number of infected individuals in each. From the 
microscopic $P\left(\vec{m},t\right) $, a distribution in the $N_{1}\times N_{1}$ square of 
integers can be defined
 \begin{align}
 \begin{split}
 &\mathcal{P}\left(n_{1},n_{2},t\right) \equiv \sum_{\left\{\vec{m}\right\}}
 \delta \left(n_{1}-\sum_{i\in \left[ 1\right]}m_{i}\right) \times\\
 &\times\delta\left(n_{2}-\sum_{i\in \left[ 2\right]}m_{i}\right) P\left(\vec{m},t\right),
 \end{split}
 \end{align}
 where $i\in \left[ g\right] $ means the individuals in subgroup $g$. After long times, 
this settles into a stationary distribution $\mathcal{P}^{\ast}\left(n_{1},n_{2}\right)$. 
Since $P^{\ast}\left(\vec{m}\right)$ is independent of the details of $\vec{m}$, 
$\mathcal{P}^{\ast}\left(n_{1},n_{2}\right)$ can be computed readily. Deep in the active 
phase, we expect it to be quite ordinary, well approximated by a Gaussian peaked around 
$(N_1,N_2) (1-1/\alpha N)$. On the other hand, $K^{\ast}$ does depend on the details of the 
partition, through $a_{j}^{i}$. The analogy between $K^{\ast}$ and the mass current leads 
us to consider a `probability angular momentum.' Associated with the stochastic time trace 
of $\left(n_{1},n_{2}\right)$ in the $N_{1}\times N_{1}$ square, such an angular momentum 
has only one component, which we will denote by $\mathcal{L}$. Furthermore, since our model 
is defined by discrete time steps, the classical velocity in $\vec{L}$ will be replaced by 
the difference $\left(n_{1}^{\prime}-n_{1},n_{2}^{\prime}-n_{2}\right)$, where 
$\left(n_{1}^{\prime},n_{2}^{\prime}\right)$ are the numbers one step later. Thus,
 \begin{align}
 &\mathcal{L}\equiv \left(n_{1},n_{2}\right) \times
 \left(n_{1}^{\prime}-n_{1},n_{2}^{\prime}-n_{2}\right)=n_{1}n_{2}^{\prime}-n_{2}n_{1}^{\prime},
 \end{align}
 while
 \begin{align}
 &\left\langle \mathcal{L}\right\rangle =\sum_{\left\{\vec{m},\vec{m}^{\prime}\right\}}
 \left(n_{1}n_{2}^{\prime}-n_{2}n_{1}^{\prime}\right)
 K^{\ast}\left(\vec{m}\rightarrow \vec{m}^{\prime}\right).
 \end{align}
 Indeed, we can venture further, using finite time differences instead of single steps:
 \begin{align}
 &\left\langle \mathcal{L}_{t}\right\rangle \equiv
 \sum_{\left\{\vec{m},\vec{m}^{\prime}\right\}}
 \left(n_{1}n_{2}^{\prime}-n_{2}n_{1}^{\prime}\right)
 Q^{\ast}\left(\vec{m}^{\prime},t;\vec{m},0\right),
 \end{align}
 where $Q^{\ast}\left(\vec{m}^{\prime},t;\vec{m},0\right) $ is the joint probability for 
finding the system in microstate $\vec{m}$ at time $0$ and in $\vec{m}^{\prime}$ $t$ steps 
later (in the NESS). Formally, $Q^{\ast}$ is given by iterating Equation (\ref{P=LP}) $t$ 
times, while $\left\langle \mathcal{L}_{t}\right\rangle$ 
is recognizable as the antisymmetric part of a (certain 
combination of a) two point, time-dependent correlations, i.e., $\left\langle 
m_{i}\left(t\right) m_{j}\left(0\right) \right\rangle $ in other common notations. In 
practice, writing down these expressions is facile, but computing them analytically is 
non-trivial and beyond the scope of this paper. Instead, we will turn to Monte Carlo 
simulations to study their properties.

\section{Simulation studies}\label{sec:sim}

Although we have some key exact results, finding expectations of macroscopic quantities is 
not feasible in general. For example, though Lenz had the explicit microscopic distribution 
for an Ising model in the 20's, two decades passed before a ferromagnetic transition is 
shown to exist (in two dimensions). In our SIS model, despite both $P^{\ast}$ and 
$K^{\ast}$ being explicitly known, many observables -- especially those associated with 
non-equilibrium statistical mechanics -- cannot be computed exactly. 
Though we expect these quantities can be well described by mean-field approximations, 
we will rely on computer simulations here. 

Specifically, we will focus on two extreme cases of
the system: $\sigma=0$ and $1$, corresponding to an undirected all-to-all network and a
directed network in which every node has $\ell$ in- and out-degrees, respectively. 
Since the connectivity differs by a factor of $2$, while the individual infection 
probabilities differ by $1/2$, the overall characteristics of the epidemic are 
indistinguishable and it is meaningful to compare the two systems. In particular, 
as we have shown in Section~\ref{sec:MEDB}, the former settles into an `equilibrium' 
system while the latter becomes a NESS. In the rest of the section we show 
simulation results, using $\alpha (N-1)= 1.6$  (corresponding to an active epidemic, 
with a level of $\thicksim 40\%$), which highlight their similarities and differences. 
Before we discuss studies with sizable $N$'s, let us present $P^{\ast}$ and $K^{\ast}$ 
for a very small system, just to verify that simulations indeed generate exact results.

 \begin{figure}[tbp]
 \subfigure[]{\includegraphics{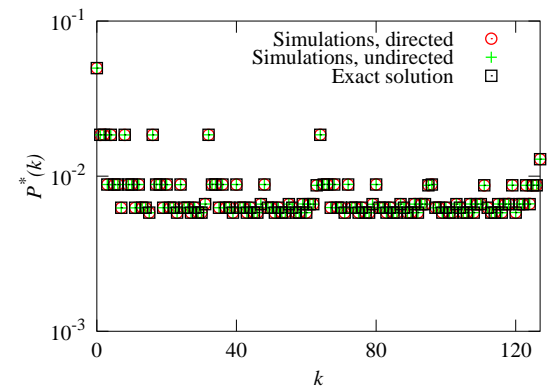}\label{fig:PuPd}}
 \subfigure[]{\includegraphics{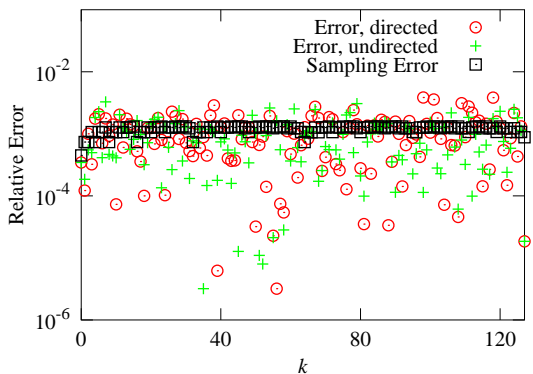}\label{fig:PuPdErr}}\newline
 \caption{(Color online) Comparison of the simulation results of the two systems with the exact solution, 
Eq.~(\ref{ExactSol}). There are $N=7$ nodes in both the directed and undirected systems, 
set at $r=1$ and $\theta \approx0.267$. 
\subref{fig:PuPd} For each value of $k$, simulation data for both systems and the 
exact results coincide. \subref{fig:PuPdErr} The relative error, 
$|1-P^{\ast}_{\text{simulations}}(k)/P^{\ast}_{\text{exact}}(k)|$, 
compared with the expected sampling error, $1/\sqrt{f_k}$, $f_k$ being the 
frequency we observe the system being in microstate $k$ during the run.}
 \label{fig:P}
 \end{figure}

 \begin{figure}[tbp]
 \subfigure[]{\includegraphics{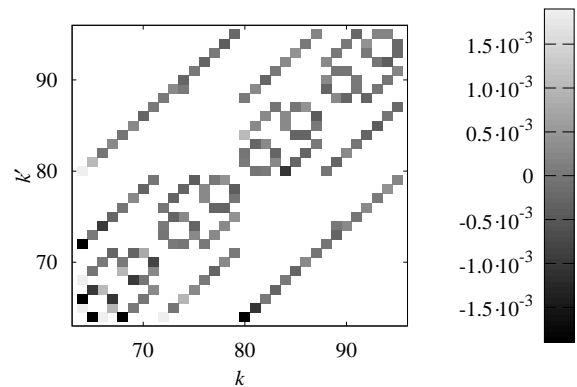}\label{fig:Fd}}
 \subfigure[]{\includegraphics{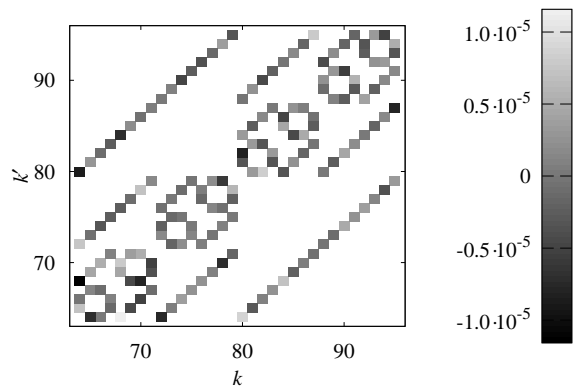}\label{fig:Fu}}\newline
 \caption{Probability currents, $K^{\ast}(k,k')$, measured in simulations for 
the~\subref{fig:Fd} directed and~\subref{fig:Fu} undirected systems. Only a portion 
of all the $128 \times 128$ currents are shown. Note the difference in the scales
for the two cases.} \label{fig:F}
 \end{figure}

 \subsection{Results for the microscopic $P^{\ast}$ and $K^{\ast}$ in a system with 
$N=7$}\label{sec:compare}

\begin{figure}[tbp]
 \subfigure[]{\includegraphics{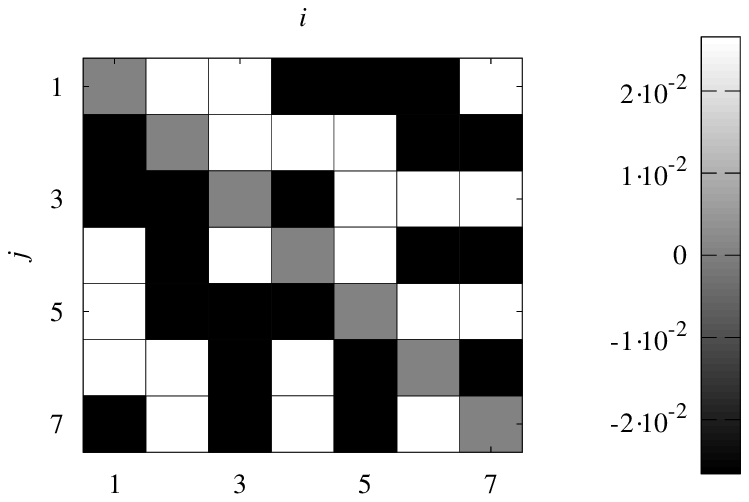}\label{fig:nFD}}
 \subfigure[]{\includegraphics{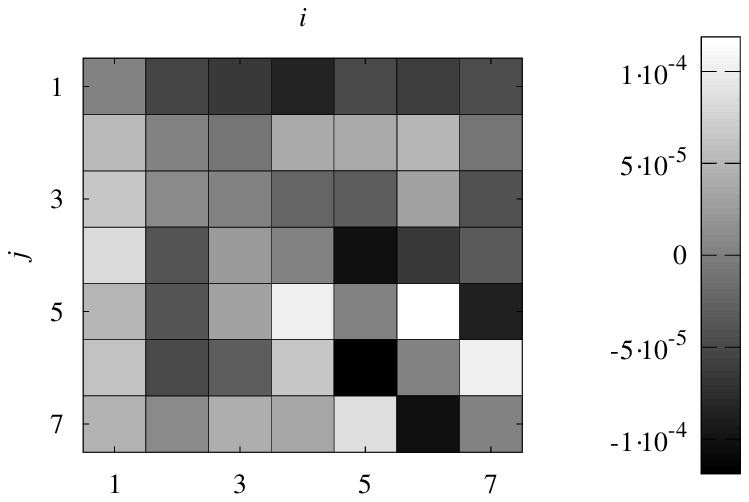}\label{fig:nFU}}\newline
 \subfigure[]{\includegraphics{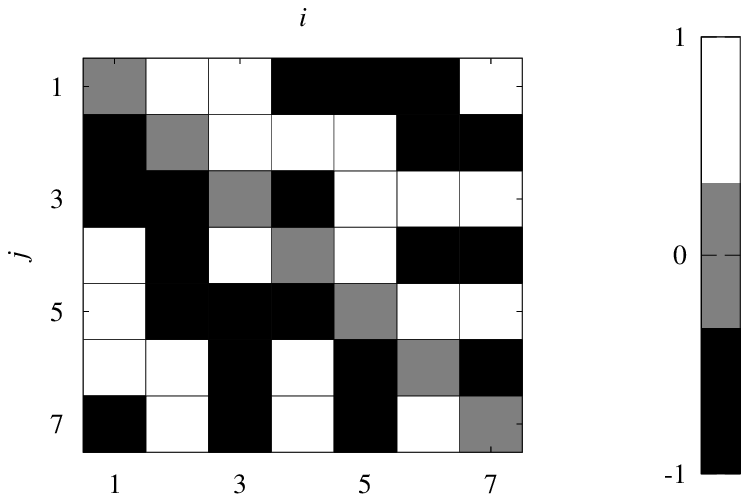}\label{fig:D}}\newline
 \caption{Coarse-grained vorticity around the $i$-$j$ plaquette in a directed
\subref{fig:nFD} and an undirected \subref{fig:nFU} network. \subref{fig:D}
 Adjacency matrix in our directed network.} \label{fig:nF}
 \end{figure}

 If we wish to compare the two approaches for these microscopic distributions, we are 
severely restricted, given that there are $2^{N}$ configurations. While $N=3$ is obviously 
trivial, we also find a special aspect to all $N=5$ systems satisfying the 
$\Sigma_{j}a_{i}^{j}=0$ constraint. Namely, the nodes can always be permuted so that their 
connectivities are identical and all graphs are circulant. At $N=7$, it is possible to 
construct several distinct classes of networks with $\Sigma_{j}a_{i}^{j}=0$. In 
Appendix~\ref{app:NC}, we provide the full algorithm for constructing a general, random 
network of this type. Returning to our particular $N=7$ system, we label the $128$ 
configurations, $\left\{m_{1},...,m_{7}\right\}$, by its binary code (e.g., 
$\left\{1,0,1,0,0,1,0\right\} \Longrightarrow k\equiv \sum_{i=1}^{N} m_i 2^{i-1}=37$). The 
specific $a_{i}^{j}$ chosen is displayed in Figure~\ref{fig:D} and we perform Monte Carlo 
simulations with the rules specified in Section~\ref{sec:model}. Typically, we discard the 
first $10^{3}$ MCS to ensure the system has settled into stationary states. Thereafter, we 
typically take measurements for the next $10^{6}$ MCS.

 First, as shown in Figure~\ref{fig:P}, simulations confirmed that the 
microscopic stationary distributions $P^{\ast}$ for both systems are 
(statistically) identical. By contrast, we display in Figures~\ref{fig:Fd} 
and~\ref{fig:Fu} the dramatic differences between the two steady state currents, 
$K^{\ast}\left(\vec{m}^{\prime}\rightarrow \vec{m}\right)$. The units correspond to the 
fraction of the time the system makes the transition $\vec{m}^{\prime}\rightarrow \vec{m}$ 
\textit{minus} the fraction of $\vec{m}\rightarrow \vec{m}^{\prime}$. Since there is no 
connection between many pairs of $\left\{\vec{m}\right\}$'s, we have illustrated the 
$2^{7}\times 2^{7}$ $K^{\ast}$-`matrix' by showing only a small section: 
$k,k^{\prime}\in\left[ 64,95\right]$. Note that, in the $\sigma=0$ system, the averages are 
consistent with zero, while the values shown are more indicative of noise. As typical 
deviations in a sampling distribution of the $K$’s, we fully expect these values to 
decrease with the length of the simulation run. In stark contrast, these averages are 
clearly non-trivial for the $\sigma \ne 0$ system, as we expect them to approach constants 
as the run time increases. Not surprisingly, these values are (statistically) the same as 
those predicted in Equation~(\ref{K*}).

Finally, at this microscopic level, we can compare the $21$ coarse-grained vorticities, 
$\omega_{ij|cg}^{\ast}$. Similar to those for $K^{\ast}$, Figure~\ref{fig:nF} shows that simulations confirm 
the theoretical results (Equations~(\ref{omega-cg}),~(\ref{eq:omegahat})). In particular, the similarity between Figures~\ref{fig:nFD} 
and~\ref{fig:D} is unmistakably clear.

\subsection{Simulation results for $N=O\left(100\right)$}\label{sec:largeN}

\begin{figure}[tbp]
 \includegraphics{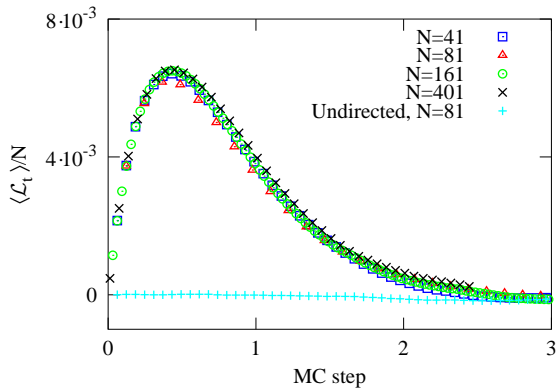}
 \caption{(Color online) Scaled plot of $\left\langle \mathcal{L}_{t}\right\rangle/N$, showing good data 
collapse for four cases: $N=41,81,161,401$. The unit of time here is MCS.}\label{fig:L}
 \end{figure}

 Lastly, we turn to more macroscopic quantities, such as $\mathcal{L}$. Intuitively, we 
expect that the effects of detailed balance violation will be maximal if \textit{all} the 
links between the two subgroups are oriented in the same direction. Due to the 
constraint $\Sigma $ $a_{i}^{j}=0$, such subgroups cannot be too large. We first performed 
simulations with $N=81$, $\alpha=0.02$, $\sigma =1$, and $N_{g}=20$ with all cross links 
between the subgroups directed from $1$ to $2$. After discarding $10^{6}$ steps ($\thicksim 
10^{4}$ MCS), we collected $n_{1,2}$ for $T\equiv 10^{8}$ steps and constructed the time 
average
 \begin{align}
 \frac{1}{T-t}\sum_{\tau =0}^{T-t}\left[ n_{1}\left( \tau \right) n_{2}\left(\tau +t\right) 
-n_{2}\left( \tau \right) n_{1}\left( \tau +t\right) \right]
 \end{align}
 as a measure for $\left\langle \mathcal{L}_{t}\right\rangle $. As a comparison, we also 
obtained similar results for the undirected case ($\sigma =0$). As in the $N=7$ 
simulations, Figure~\ref{fig:L} shows the dramatic difference in 
$\left\langle\mathcal{L}_{t}\right\rangle$ between the two models. The most prominent 
feature is that $\left\langle \mathcal{L}_{t}\right\rangle$ is positive. The same 
intuitive picture offered above for this sign can be restated here. Since the links all 
direct from $1$ to $2$, we may expect that a fluctuation in $n_{1}$ (say, increase) will 
lead, in the next few steps, to more infected individuals in subgroup $2$. By contrast, 
outbreaks in the latter do not affect those in subgroup $1$.

 At present, we have no quantitative explanation for the other notable feature: the rise 
and fall of $\left\langle \mathcal{L}_{t}\right\rangle$ as a function of $t$. Nevertheless, 
we may consider the following argument. Since the data is plotted against time 
\textit{steps}, we can reasonably expect that it take $O\left( N_{g}\right) $ steps before 
correlations associated with the collective behavior of the group is built up. On the other 
hand, the system is far from being critical, so that we may expect finite correlation 
times, which would lead to decays at large $t$. To see if these notions are worth pursuing, 
we carry out a simple scaling analysis, using $N_{g}=10,20,80,100$ in populations with 
$N=41,81,161,401$ and correspondingly modified $\alpha =0.04,0.02,0.01,0.004$. As 
Figure~\ref{fig:L} shows, we find excellent data collapse when 
$\left\langle\mathcal{L}_{t}\right\rangle/N$ is plotted against $t/N$ (i.e., MCS). The 
scaling $\left\langle\mathcal{L}_{t}\right\rangle \thicksim N$ can be argued as follows. 
Though we expect each $n_{\alpha}$ to scale with $N$, the quantities which enters into 
$\mathcal{L}$ are actually deviations from $\left\langle n_{\alpha }\right\rangle$. If we 
naively assume that the deviations scale as $\sqrt{N}$, then we arrive at 
$\left\langle\mathcal{L}_{t}\right\rangle \thicksim N$. 
Work is in progress on both the simulation and the theoretical fronts, to draw reliable
conclusions and to achieve an in-depth understanding of these phenomena.

\begin{figure}[tbp]
 \subfigure[]{\includegraphics{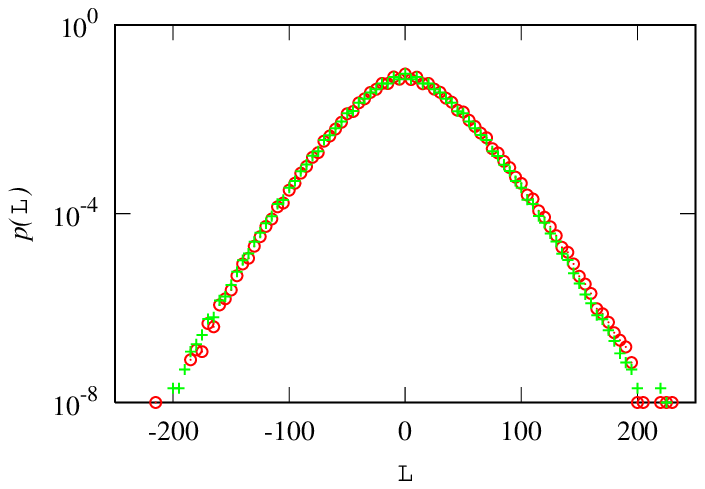}\label{fig:PPA}}
 \subfigure[]{\includegraphics{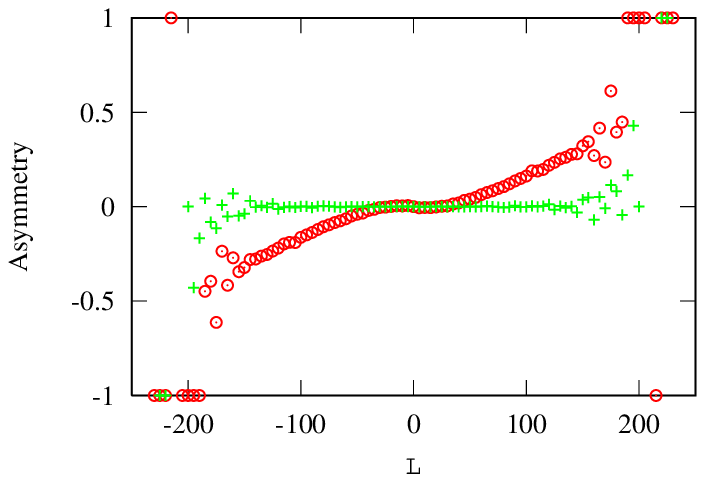}\label{fig:PPB}}
 \caption{(Color online) \subref{fig:PPA} Distributions of $\mathcal{L}$, $p\left(\mathcal{L}\right)$, 
obtained from histograms of observed values in a run of $10^{8}$ steps, for the undirected 
network ($\sigma=0$, green pluses) and the directed one ($\sigma=1$, red open circles). 
\subref{fig:PPB}Asymmetry in the distributions, defined as 
$\left[p\left(\mathcal{L}\right)-p\left(-\mathcal{L}\right)\right]/\left[p\left(\mathcal{L}\right)+p\left(-\mathcal{L}\right)\right]$, 
highlighting the different behaviors associated with the undirected network ($\sigma=0$, 
green pluses) and the directed one ($\sigma=1$, red open circles).} \label{fig:PP}
 \end{figure}

 As a final note, we present relevant data concerning the \textit{fluctuations} in 
$\mathcal{L}$, since a valid question could be raised concerning the standard deviation 
associated with the observed averages. To appreciate better such issues, let us first 
illustrate with one particular case -- $N=81, \alpha=0.02, t=40$ (corresponding to the peak 
in Figure~\ref{fig:L}), by displaying the full distributions of the observed 
$\mathcal{L}$'s, $p\left(\mathcal{L}\right)$, for both the $\sigma=0$ vs.~$1$ models. Since 
$\left\vert\mathcal{L}\right\vert\leq 20^{2}$, the range shown here is reasonable. While 
the two curves in Figure~\ref{fig:PPA} are quite broad and give the impression of being 
indistinguishable, a plot of the asymmetry in Figure~\ref{fig:PPB} clearly displays the 
difference. As a result of this asymmetry, $\left\langle \mathcal{L}_{40}\right\rangle 
\cong 0.512$ in the NESS case. By contrast, it is consistent with zero ($\thicksim 
10^{-4}$) for the undirected network. Quantitatively, the standard deviations for the $\sigma 
=\left( 0,1\right) $ cases are, respectively, approximately $\left(27.6,27.8\right)$, with 
skewness $\left(10^{-4},0.0597\right)$ and kurtosis $\left(0.704,0.6961\right)$. Clearly 
not Gaussians, these distributions deserve to be studied in further detail. Similarly, 
there appears to be interesting features in the asymmetry plot. We should pursue them and 
ask if their origin is merely a chance fluctuation or some systematic intriguing physics.

\section{Summary and Outlook}

 We study a simple SIS model of epidemics on a complete graph with infection rates that 
interpolate between symmetric ($\sigma =0$) and fully antisymmetric ($\sigma =1$). In the 
language of graphs, these correspond to undirected and directed ones, respectively, the 
latter associated with an antisymmetric (part of the) adjacency matrix $a_{i}^{j}$. To make 
comparisons between models with different $\sigma $ meaningful, we impose a restriction: 
$\Sigma_i $ $a_{i}^{j}=0$, i.e., every node has the same number of in- and out-degrees 
when $\sigma =1$. With 
relatively high infection rates (and a small reinfection probability to avoid being trapped 
in the absorbing state), the system settles into an active state, which is an equilibrium 
stationary state or a non-equilibrium one, respectively. Solving the master equation 
exactly, we find that the stationary distribution, $P^{\ast }$, is \textit{ independent} of 
$\sigma $. Such a result is reminiscent of the one in the asymmetric simple exclusion 
process~\cite{Spitzer70}, in which $P^{\ast}\propto 1$, regardless of the strength of the 
bias. Thus, static properties, such as phase transitions, critical behavior and equal time 
correlations, will also be independent of $\sigma$.

 On the other hand, the dynamics of a $\sigma >0$ system violates detailed balance, so that 
non-vanishing steady (probability) currents will be present. In the stationary state, these 
must form closed loops, as in mangetostatics. Their consequences will be observable only 
when dynamic quantities (e.g., unequal time correlations) are measured. At the microscopic 
level, these current loops form vortices around a plaquette associated with a pair of 
nodes: $\left(i,j\right) $. The vorticities, also found exactly, are proportional 
to, as expected, $\sigma a_{i}^{j}$. Physically, they correspond to the frequency of 
cyclic infection-recovery behavior: in ($SS\rightarrow SI\rightarrow II\rightarrow 
IS\rightarrow SS$) \textit{vs.} the reverse loop. At the macroscopic level, we can consider 
two groups of individuals and the numbers of the infected: $\left( n_{1},n_{2}\right) $. One 
consequence of non-zero probability currents is that, in general, trajectories in the 
$n_{1}$-$n_{2}$ plane are more likely to circulate one way rather than the other. We focus 
on a particular quantity, $\mathcal{L}$, which is the analog of angular momentum in 
classical mechanics and being studied in the context of the climate science~\cite{OHC}. 
Dubbed the `probability angular momentum,' it is simply the antisymmetric part of an 
unequal-time correlation between two quantities. Illustrating with a specific example, 
$\left\langle n_{1}\left(0\right)n_{2}\left(t\right)-n_{2}\left(0\right)n_{1}\left(t\right)\right\rangle$ is found to display interesting properties. Though the 
qualitative aspects are expected, much of the quantitative features remains to be analyzed.

 Naturally, our study here raises many interesting questions, from those related to SIS 
models to a wider spectrum of systems in non-equilibrium steady states. For our SIS model, 
we fully expect that, deep in the active phase, the fluctuations and correlations can be 
well approximated by a linear Langevin equation, leading to Gaussian (but non-equilibrium) 
distributions~\cite{Lax,LGM1,LGM2,ZS07}. The associated currents are well 
understood~\cite{ZS07} and distributions for collective quantities like 
$p\left(\mathcal{L}\right)$ can then be computed~\cite{OHC}.

Beyond our simple system with all-to-all connections, there are many SIS
models, cast in the context of a variety of networks (e.g., square periodic
lattice)~\cite{Moore2000,Pastor-Satorras2001,Keeling2005,Barrat2008}.
Further, to model realistic epidemics, SIS is too simplistic. In more
complex models, it is also very likely that their evolution violate detailed
balance, so that persistent probability currents should be present in those
steady (or quasi-stationary) states. We are not aware of any studies on
observable consequences of these currents and believe that such pursuits can
yield new insights into both cylclic behavior in a quasi-stationary ongoing
epidemic and the variety of paths to its extinction. We expect the results
presented here to provide some guidance in the search for novel
manifestations of probability currents. 

In a wider context, since probability currents necessarily persist in NESS 
\cite{ZS07}, the study of their observable manifestations is of some
importance. The range of these manifestations in nature is enormous, from
convection cells of all varieties and sizes (Raleigh-Benard,
Kelvin-Helmholtz) to energy/matter fluxes \textit{through} all living
organisms. The relationship between microscopic probability currents and
such macroscopic phenomena has been explored in, e.g., \cite{ZS07}. Two
intriguing possibilities exist. One is that, under coarse-graining, the
effects of these currents become less and less relevant (in the
renormalization group sense). There are few investigations on how such
renormalization group flows, despite the importance of understanding this
class of systems. To study the other possibility -- effects surviving
coarse-graining -- is clearly more urgent, since macroscopic currents are
essential for life and ubiquitous in nature. Of course, our distant goal
lies far beyond the models of epidemics considered here. It is to develop an
overarching framework to characterize such behavior for all stochastic
processes which allow the system to settle into non-equilibrium steady
states. In such a framework, probability distributions of currents will play
a central role, just as the probability distributions of configurations is
central to equilibrium statistical mechanics.

 \begin{acknowledgements}
 We acknowledge fruitful discussions with B. Fox-Kemper, 
D. Mandal, B. Schmittmann, Z. Toroczkai, and J.B. Weiss. This research 
is supported in part by the US National Science 
Foundation through grants DMR-1244666 and DOS-1245944.
 \end{acknowledgements}

 \appendix

 \section{Detailed Balance and Kolmogorov Criterion}\label{DB+KC}

 Since all loops in configuration space can be regarded as sums of `elementary' loops, each 
winding around a face (plaquette), this criterion can be checked by studying the 
product of $\Omega $'s around an elementary loop. Thus, we consider the sequence
 \begin{align}
 &\left(m_{i},m_{j}\right) =\left(0,0\right) \rightarrow \left(1,0\right)
 \rightarrow \left(1,1\right) \rightarrow \left(0,1\right), \label{DBKC1}
 \end{align}
 with all other entries ($m_{k\neq i,j}$) held fixed. For simplicity, we use only these two 
$m$'s as shorthand to stand for the four configurations. Thus, if we define $n_{0}\equiv\Sigma_{k\neq i,j}m_{k}$, and $n_{1,2}=n_{0}+1,2$, the sequence of $n\left(\vec{m}\right)$'s and $\rho $'s are
 \begin{align}
 &n_{0} \rightarrow n_{1}\rightarrow n_{2}\rightarrow n_{1}\rightarrow n_{0},\\
 &\rho_{0} \rightarrow \rho_{1}\rightarrow \rho_{2}\rightarrow \rho_{1}\rightarrow 
\rho_{0},
 \end{align}
 where $\rho_{\alpha}\equiv 1/\left[ n_{\alpha}\left\{r+\theta 
\left(N-n_{\alpha}\right)\right\} \right]$ is just a short hand for $\rho_{n_{\alpha}}$. 
With this notation, the associated product of the transition probabilities is 
$\Pi \equiv \Omega_{i}\left(0,0\right)\Omega_{j}\left(1,0\right) \Omega_{i}\left(1,1\right) \Omega_{j}\left(0,1\right) $, i.e.,
 \begin{align}
 &\Pi =\rho_{0}\gamma_{i}\left(0,0\right) \rho_{1}\gamma_{j}\left(1,0\right) 
\rho_{2}n_{2}r\rho_{1}n_{1}r,\label{product}
 \end{align}
 For the loop in reverse, the product is $\Pi_{R}\equiv \Omega_{i}\left(0,0\right) 
\Omega_{j}\left(0,1\right) \Omega_{i}\left(1,1\right) \Omega_{j}\left(1,0\right) $, i.e.,
 \begin{align}
 &\Pi_{R}=\rho_{0}\gamma_{j}\left(0,0\right) \rho_{1}\gamma_{i}\left(0,1\right) 
\rho_{2}n_{2}r\rho_{1}n_{1}r.
 \end{align}
 The Kolmogorov criterion, $\Pi \overset{?}{=}\Pi_{R}$, reduces to
 \begin{align}
 &\gamma_{i}\left(0,0\right) \gamma_{j}\left(1,0\right) 
\overset{?}{=}\gamma_{j}\left(0,0\right) \gamma_{i}\left(0,1\right).
 \end{align}
 Using Eqn.~(\ref{gamma2}), this test becomes
 \begin{align}
 \begin{split}
 &\left[ n_{0}+\sigma \kappa_{i}\left(0,0\right) \right] \left[n_{1}+\sigma 
\kappa_{j}\left(1,0\right) \right] \overset{?}{=}\\
 & \overset{?}{=}\left[n_{0}+\sigma \kappa_{j}\left(0,0\right) \right] \left[ n_{1}+\sigma 
\kappa_{i}\left(0,1\right) \right],
 \end{split}
 \end{align}
 or
 \begin{align}
 \begin{split}
 &\sigma \left\{n_{0}\left[ \kappa_{j}\left(1,0\right) -\kappa_{i}\left(0,1\right) 
\right]\right.+\\ 
 &+\left. n_{1}\left[ \kappa_{i}\left(0,0\right)-\kappa_{j}\left(0,0\right) \right] 
\right\} +\\
 &+\sigma ^{2}\left[ \kappa_{i}\left(0,0\right) \kappa_{j}\left(1,0\right) 
-\kappa_{j}\left(0,0\right) \kappa_{i}\left(0,1\right) \right]\overset{?}{=}0,
 \end{split}
 \end{align}
 where
 \begin{align}
 \begin{split}
 &\kappa_{i}\left(\mu ,\mu^{\prime}\right)=\sum_{k}a_{i}^{k}m_{k}+a_{i}^{j}\mu^{\prime},\\
 &\kappa_{j}\left(\mu ,\mu^{\prime}\right)=\sum_{k}a_{j}^{k}m_{k}+a_{j}^{i}\mu.
 \end{split}
 \end{align}

 Clearly, the equality can fail provided $\sigma >0$ and so, the differences above do not 
vanish for a general $\vec{m}$. Note however, that it does vanish with $\sigma $, which 
indicates that, not surprisingly, SIS on a complete, undirected graph settles into an 
equilibrium state.

 \section{Normalization Factor}\label{norm}

 To compute the sum in Eqn.~(\ref{P0}), we consider
 \begin{align}
 &S_{N}\left(\alpha \right) \equiv 
\sum_{n=0}^{N}\binom{N}{n}n!\alpha^{n}=\int_{0}^{\infty}\text{d}xe^{-x}\left(1+\alpha 
x\right)^{N},\label{eq:SN}
 \end{align}
 where $n!=\int e^{-x}x^{n}$ is used and the sum is performed first. Changing the 
integration variable to $y=x+1/\alpha $, this becomes
 \begin{align}
 \begin{split}
 &S_{N}\left(\alpha \right) =\alpha^{N}e^{1/\alpha}\int_{1/\alpha}^{\infty}\text{d}y 
e^{-y}y^{N}=\\
 &=\alpha^{N}e^{1/\alpha}\Gamma \left(N+1,1/\alpha\right),
 \end{split}
 \end{align}
 where $\Gamma $ is the upper incomplete gamma function. Thus,~(\ref{P0}) can be written as
 \begin{align}
 &\frac{1}{P_{0}}=\frac{1}{\alpha 
N}\sum_{n=1}^{N}\binom{N}{n}n!\alpha^{n}+1+\sum_{n=1}^{N}\binom{N}{n}n!\alpha 
^{n}\left(\frac{N-n}{N}\right).
 \end{align}
 But,
 \begin{align}
 \begin{split}
 &\frac{1}{\alpha N}\sum_{n=1}^{N}\binom{N}{n}n!\alpha 
^{n}=\sum_{n=1}^{N}\frac{\left(N-1\right) !}{\left(N-n\right) !}\alpha ^{n-1}=\\
 &=\sum_{m=0}^{N-1}\frac{\left(N-1\right)!}{\left(N-1-m\right)!}\alpha^{m},
 \end{split}
 \end{align}
 while
 \begin{align}
 &1+\sum_{n=1}^{N-1}\frac{\left(N-1\right) !}{\left(N-1-n\right) 
!}\alpha^{n}=\sum_{n=0}^{N-1}\frac{\left(N-1\right) !}{\left(N-1-n\right) !}\alpha^{n},
 \end{align}
 so that both are $S_{N-1}\left(\alpha \right) $. Thus, we arrive at a compact expression:
 \begin{align}
 &P_{0}=\frac{1}{2 S_{N-1}(\alpha)}.
\label{P0-final}
 \end{align}

 \section{Coarse-grained vorticity around a plaquette}\label{vorticity}

 Consider a pair of individuals, $i$ and $j$, providing four states, $\left( 
m_{i},m_{j}\right) $, and the net currents around the plaquette (as in Eqn. \ref{DBKC1}). 
Defining $\hat{m}$ as $\vec{m}$ \textit{without} the pair $\left( m_{i},m_{j}\right)$, 
then the persistent currents around the loop are
 \begin{align}
 \begin{split}
 &K^{\ast}\left(\left\{0,0,\hat{m}\right\} \rightarrow \left\{1,0,\hat{m}\right\} \right)=\\
 & =\sigma\left(P_{0}/N\right) \alpha ^{\nu}\left(\nu -1\right)!\sum_{k\neq i,j}a_{i}^{k}m_{k},
 \end{split}\\
 \begin{split}
 &K^{\ast}\left(\left\{0,1,\hat{m}\right\} \rightarrow \left\{1,1,\hat{m}\right\}\right)=\\
 &=\sigma \left(P_{0}/N\right) \alpha ^{\nu +1}\nu !\left[\sum_{k\neq i,j}a_{i}^{k}m_{k}+a_{i}^{j}\right],
 \end{split}\\
 \begin{split}
 &K^{\ast}\left(\left\{1,0,\hat{m}\right\} \rightarrow \left\{1,1,\hat{m}\right\} \right)=\\
 &=\sigma\left(P_{0}/N\right) \alpha ^{\nu +1}\nu !\left[\sum_{k\neq i,j}a_{j}^{k}m_{k}+a_{j}^{i}\right],
 \end{split}\\
 \begin{split}
 &K^{\ast}\left(\left\{0,0,\hat{m}\right\} \rightarrow \left\{0,1,\hat{m}\right\} \right)=\\
 &=\sigma\left(P_{0}/N\right) \alpha^{\nu}\left(\nu -1\right)!\sum_{k\neq i,j}a_{j}^{k}m_{k},
 \end{split}
 \end{align}
 where $\nu \equiv n\left( 0,0,\hat{m}\right) $ is the number of infected in $\hat{m}$, and 
must be positive here. The $\nu =0$ case is special, as $K^{\ast }\left( \vec{0}\rightarrow 
\left\{ 1,0,...,0\right\} \right) =0$.

Now, the vorticity around this plaquette is
 \begin{align}
 \begin{split}
 &\omega_{ij}^{\ast} \equiv K\left(\left\{0,0,\hat{m}\right\} \rightarrow 
\left\{1,0,\hat{m}\right\} \right) +\\
 &+K\left(\left\{1,0,\hat{m}\right\} \rightarrow \left\{1,1,\hat{m}\right\} \right)- \\
 &-K\left(\left\{0,1,\hat{m}\right\} \rightarrow \left\{1,1,\hat{m}\right\} \right)-\\ 
 &-K\left(\left\{0,0,\hat{m}\right\} \rightarrow \left\{0,1,\hat{m}\right\} \right).
 \end{split}
 \end{align}
Thus,
\begin{align*}
\begin{split}
& \omega _{ij}^{\ast }=\sigma \left( P_{0}/N\right) \alpha ^{\nu }\left( \nu
-1\right) !\times  \\
& \times \left[ (\alpha \nu -1)\sum_{k\neq i,j}(a_{j}^{k}-a_{i}^{k})m_{k}
+2\nu\alpha a_{j}^{i}\right],
\end{split}
\end{align*}
for $\nu>0$, and simply $2\sigma\alpha a_{j}^{i}\left( P_{0}/N\right) $ for $\nu=0$. 
To proceed further, let us define the `coarse-grained' vorticity,
\begin{align}
\omega _{ij|cg}^{\ast }\equiv \sum_{\left\{ \hat{m}\right\} }\omega_{ij}^{\ast }=
\hat{\omega}a_{j}^{i},
\end{align}
and find $\hat{\omega}$. First, note that, for any $k$,
\begin{align}
\begin{split}
&\sum_{\left\{ \hat{m}\right\} }m_{k}\delta \left( \nu -\sum_{\ell \neq i,j}m_{\ell }\right)=\\
&=\sum_{\left\{ \hat{m}\right\} }\frac{\Sigma_{k}m_{k}}{N-2}\delta \left( ...\right)=\frac{\nu }{N-2}\binom{N-2}{\nu}.
\end{split}
\end{align}
Next, note 
 \begin{align}
 &0=\sum_{k}a_{j}^{k}=\sum_{k\neq i,j}a_{j}^{k}+a_{j}^{i},
 \end{align}
 and
 \begin{align}
 &\sum_{\left\{ \hat{m}\right\} }=\sum_{\nu }\sum_{\left\{ \hat{m}\right\}}
 \delta \left( \nu -\sum_{\ell \neq i,j}m_{\ell }\right). 
 \end{align}
 Thus,
 \begin{align}
 \begin{split}
 &\omega _{ij|cg}^{\ast }=\frac{2\sigma a_{j}^{i}P_{0}}{N} \Bigg [ \alpha+\\
 &+\sum_{\nu >0}\binom{N-2}{\nu }\alpha ^{\nu }\nu! \left[ \alpha +\frac{1-\alpha \nu }{N-2}\right] \Bigg].
 \end{split}
 \end{align}
 Rewriting
 \begin{align}
 &\alpha +\frac{1-\alpha \nu }{N-2}=\frac{1}{N-2}+\alpha \frac{N-2-v}{N-2}
 \end{align}
 and combining the last term:
 \begin{align}
 \begin{split}
 &\alpha+\sum_{\nu >0}\binom{N-2}{\nu }\alpha ^{\nu }\nu !\left[ \alpha \frac{N-2-v}{N-2} \right] =\\
 &=\alpha\sum_{\nu =0}^{N-3}\binom{N-3}{\nu }\alpha ^{\nu }\nu!,
 \end{split}
 \end{align}
 we find 
 \begin{align}
 \hat{\omega}=\frac{\sigma }{NS_{N-1}\left( \alpha \right) }\left( \frac{S_{N-2}(\alpha )-1}{N-2}+\alpha S_{N-3}(\alpha )\right).
 \end{align}

\section{Network Construction}\label{app:NC}

The matrix $a_{j}^{i}$ in the numerical simulations was constructed using the following 
algorithm:

\begin{enumerate} \item begin with an all-to-all connected network

\item pick a node

\item compute the current difference, $\Delta$, between the outgoing and incoming degrees, 
where an undirected edge makes no contribution to the difference

\item pick an undirected edge attached to that node

\item assign a direction to chosen edge according to the following rule: if $\Delta=1$ make 
the edge incoming, if $\Delta=-1$ make the edge outgoing, if $\Delta=0$ pick the edge 
direction at random

\item consider the node at the other end of that edge

\item repeat steps 3 through 6

\item arriving at a node with no undirected edges, randomly pick a node that still has 
undirected edges and repeat steps 2 through 7 \end{enumerate}

Thus, we trace out the entire network assigning the edge directions, making sure that the 
current incoming and outgoing degrees of the current node are equal. In the end of this 
process we obtain a network where every node has equal in and out degrees.

\bibliographystyle{apsrev} \bibliography{Shkarbib}

\end{document}